\documentclass[conference]{IEEEtran}
\IEEEoverridecommandlockouts
\usepackage{graphicx, amsmath, amssymb, subcaption, url, cite, array, amsthm}
\usepackage[ruled, linesnumbered]{algorithm2e} 
\usepackage{algorithm2e, setspace}
\usepackage[font={small}]{caption}

\begin{document}
\title{\huge Toward BCI-enabled Metaverse: A Joint Learning and Resource Allocation Approach}

\author{\IEEEauthorblockN{Nguyen Quang Hieu, Dinh Thai Hoang, Diep N. Nguyen, and Eryk Dutkiewicz\\} 
\IEEEauthorblockA{School of Electrical and Data Engineering, University of Technology Sydney, Australia}}

\maketitle
\begin{abstract}
In this paper, we propose a framework that uses Brain-Computer Interface (BCI) technology to create human-like avatars for user-driven Metaverse applications. This framework is designed to work efficiently with fast wireless connectivity and high computing demand, making it ideal for future infrastructures, e.g., 5G and beyond.
The Metaverse system uses brain signals sent through wireless channels to create intelligent digital avatars that can provide helpful recommendations and assist in user-driven applications.
To eliminate the computational burden on the user equipments, the computational tasks and resource allocation decisions are shifted to the centralized base station.
As a result, our framework involves solving a mixed decision-making and classification problem. The goal is for the base station to efficiently allocate its computing and radio resources to users, as well as classify their brain signals. To this end, we develop a hybrid training algorithm that uses the latest advancements in deep reinforcement learning to solve the problem. Our algorithm involves three deep neural networks working together to handle both decision-making and classification tasks.
Simulation results indicate that our framework can effectively manage system resources while accurately classifying users' brain signals.
\end{abstract}

\section{Introduction}
Recent years have witnessed the fast development of user-driven applications with high data rate demand and tremendous computing requirements such as virtual reality (VR) streaming and gaming \cite{saad2019}. 
The future 5G and 6G networks are envisioned to provide fast network connectivity and intelligent user-driven services for such VR applications.
Toward this convergence of faster connectivity and user-driven applications, the recently emerged Metaverse concept has attracted enormous attention from both academia and industry \cite{xu2022}.
Moreover, the development of the Metaverse has been facilitated by recent technological breakthroughs such as VR, blockchain, AI, 5G, and beyond \cite{xu2022}.  
In Metaverse, the users can interact with Metaverse environment and other users through their avatars. 
Despite the growing attention from industry and academia, Metaverse is still in its infancy. 
With virtual avatars as the human embodiments in Metaverse, the avatars can be potentially reflected user individual characters. 
Although Metaverse inherits visual components from virtual reality and augmented reality platforms, the intelligence of such avatars is still an uncovered topic.    

Toward the effort of making intelligent digital avatars for Metaverse, a research field evolves at the intersection of neuroscience and virtual reality. The idea is to make digital avatars more intelligent and more individual by using biological signals from humans. Specifically, using brain signals is one of the most promising methods with a long history in the development of Brain-Computer Interface (BCI) \cite{lotte2012}. 
In \cite{lee2022}, an imagined speech communication system toward Metaverse is proposed. The electroencephalography (EEG) signals of users are analyzed to predict the imagined words. The authors also propose a prototype for a virtual assistant avatar in a smart home as a potential Metaverse application. 
There are several approaches toward BCI-enabled VR applications in the literature. In \cite{chun2016}, the authors propose BCI-based methods for navigation tasks in a VR environment. Extension of such BCI-based methods can be virtual autonomous driving \cite{qibin2009} and adaptive VR environment rendering \cite{lotte2012}.
Although the aforementioned works achieve adequate performance for VR  applications, the step toward Metaverse which usually involves tremendous computational demand and complex interactions between multiple human-like avatars and is still a big research gap. 
For example, the wired connections between the BCI devices and computing units in conventional settings in \cite{lotte2012, lee2022}, and \cite{qibin2009} might not be always feasible and available due to coverage and mobility problems \cite{saad2019}.
Thus, the integration of BCI into cellular-based systems is a promising solution for future Metaverse applications.

To this end, in our paper, we first propose an innovative framework in which Metaverse users with integrated VR-BCI headsets can immerse VR applications while sending BCI signals toward uplink wireless channels. 
As such, the base station (BS) can create intelligent human-like avatars to further enhance user Quality-of-Experience (QoE). In particular, our proposed framework involves a mixed decision-making and classification problem.
The decision-making problem requires the resource allocation policy to be derived so that the VR delay of the users is minimized. 
The classification problem requires highly accurate predictions of brain signals for facilitating the creation of intelligent digital avatars. 
As the decision-making and classification problems are shifted to a wireless edge server or a BS, the computational burden can be significantly reduced at the user devices, i.e., VR headsets and EEG headsets, thus enabling lightweight design and feasible antenna deployment.  
However, it is very challenging to address not only the resource allocation problem but also the brain signals classification problem.
To this end, we propose a novel hybrid learning algorithm that leverages the advantages of deep reinforcement learning to optimize the resources of the system and properly predict the brain signals of the users with high accuracy.

\section{System Model}
\begin{figure}[t]
\centering
\includegraphics[width=0.55\linewidth]{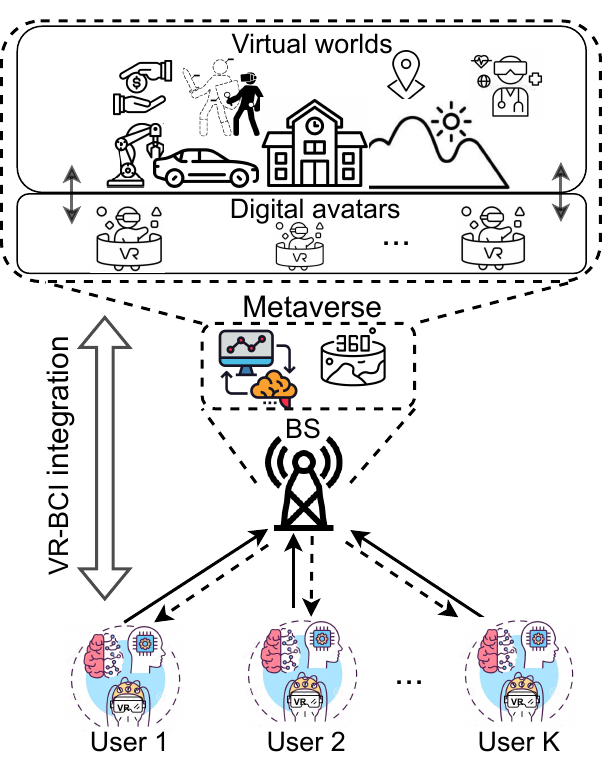}
\caption{An illustration of our system model. $K$ users equipped with integrated VR-BCI headsets are experiencing VR applications transmitted by the BS via downlink channels. At the same time, the BS collects BCI signals from the users via uplink channels for creating digital avatars to enhance user QoE.}
\label{fig:system-model}
\end{figure}

Our proposed system model is illustrated in Fig.~\ref{fig:system-model}.
The system consists of (i) a Base Station (BS) and (ii) $K$ users equipped with integrated VR-BCI headsets, e.g., Galea \cite{bernal2022}.
Each integrated VR-BCI headset can extract BCI signals from $J$ channels (i.e., corresponding to $J$ electrodes of the headset) from the user and provide VR services for the user at the same time.
The operations of the system are as follows.
\subsection{System Operation}

At time step $t$, user $k$ sends BCI signals $\mathbf{e}_k(t)$ to the BS. The BCI signals $\mathbf{e}_k(t) \in \mathbb{R}^{J \times W}$ can be represented by a $J \times W$ vector where $J$ is the number of BCI channels and $W$ is the number of BCI signals collected during the sampling interval in time step $t$. At the end of the time step $t$, the BS can collect a set of BCI signals, i.e., 
\vspace{-0.2cm}
\begin{equation}
\mathbf{e}(t) = \{\mathbf{e}_1(t), \mathbf{e}_2(t), \ldots, \mathbf{e}_K(t)\}\in \mathbb{R}^{K \times J \times W}.
\label{eq:bci-signal-vector}
\end{equation}
At the same time, the BS pre-processes VR content for users. The pre-processing process can be 360-degree video viewport rendering \cite{fernandes2016} or changing video resolution \cite{kim2017} that can enhance the user Quality-of-Experience (QoE). For example, VR sickness\footnote{VR sickness is a common issue in VR
applications in which the user’s brain receives conflicting
signals about self-movement between the physical and virtual
worlds.} or user fatigue can be detected from BCI signals as shown in \cite{chang2020}.
The BS monitors the total computing load (i.e., CPU), denoted by $u_n(t) \in (0, 1)$, of the $n$-th core that is available at the BS at time step $t$.
The information about total computing load, denoted by $\mathbf{u}(t) \in \mathbb{R}^N$, is defined by:
\vspace{-0.2cm}
\begin{equation}
\mathbf{u}(t) = \{u_1(t), u_2(t), \ldots, u_N(t)\},
\label{eq:cpu-load}
\end{equation} 
where $N$ is the number of CPUs of the BS.
The BS then analyses the current state of the system and calculates the optimal policy, i.e., computing resource allocation for VR pre-processing and radio/power resource allocation for the uplink channels in the next time step $t+1$.

To evaluate the performance of the proposed framework, we propose a QoE metric that is a combination of (i) the round-trip VR delay at the user and (ii) the accuracy of classifying the BCI signals at the BS. The round-trip VR delay is the latency between the time the user requests VR content from the BS and the time the user gets the requested VR content displayed in his/her headset. 
The accuracy of analyzing BCI signals is obtained from a predictor at the BS that can predict the actions of the users. 
We choose our metrics based on their ability to reduce VR sickness and fatigue for the users. Studies in \cite{chang2020, chun2016, fernandes2016} support this approach. Additionally, we focus on BCI signal classification because predicting user actions can help create intelligent avatars that behave like humans in Metaverse scenarios. This could include imagined speech communication \cite{lee2022}, adaptive VR environment rendering \cite{lotte2012}, and detecting anomalous states and error-related behaviors \cite{arico2017}. To formulate our QoE metric, we construct a round-trip VR delay and BCI predictor as follows.

\subsection{Round-trip VR Delay}
The round-trip VR delay consists of (i) processing latency at the BS, (ii) downlink transmission latency, and (iii) uplink transmission latency. Since most of the computation is shifted to the BS, we assume that the latency at the user headsets is negligible. Accordingly, the round-trip VR delay of user $k$ at time step $t$ is calculated by:
\vspace{-0.2cm}
\begin{equation}
D_k(t) = d_k(t) + \frac{l_k^D}{r_k^D(t)} + \frac{l_k^U}{r_k^U(t)},
\label{eq:system-latency}
\end{equation}
where $d_k(t)$ is the processing latency, e.g., pre-rendering the VR content, at the BS, $l_k^D$ and $l_k^U$ are the length of data packets to be transmitted over the downlink and uplink, respectively. $r_k^D(t)$ and $r_k^U(t)$ are the downlink and uplink data rates between the user $k$ and the BS, respectively.
The processing delay of the BS depends on the process running at the BS and the CPU capacity of the BS. In our setting, we consider that the BS can support a VR streaming service. 
At time step $t$, the BS measures its current available CPU state $u_n(t) \in \mathbf{u}(t)$.
Let $\tau_k(t) \in (0, 1)$ denote the portion of $u_n(t)$ that is used for  pre-processing VR content for user $k$.
Once $u_n(t)$ and $\tau_k(t)$ are obtained, the VR pre-processing delay for user $k$ at the BS can be calculated by:
\vspace{-0.2cm}
\begin{equation}
d_k(t) = \frac{1}{\tau_k(t) u_n(t) \upsilon},
\label{eq:processing-delay}
\end{equation}
where $\upsilon$ (Hz) is the CPU capacity, i.e., the total number of computing cycles, of the BS.
The uplink data rate for user $k$ is defined as follows:
\vspace{-0.2cm}
\begin{equation}
r_k^U(t) = \sum_{m \in \mathcal{M}} B^U \rho_{k,m}(t) \log_2\Big(1 + \frac{p_{k}(t) h_{k}(t)}{I_m + B^U N_0}\Big),
\label{eq:uplink-rate}
\end{equation}
where $\mathcal{M}$ is the set of radio resource blocks, $p_{k}(t)$ is the transmit power of the user $k$, and $h_{k}$ is the time-varying channel between the BS and user $k$. $\rho_{k,m}(t) \in \{0, 1\}$ is the resource block allocation variable. $\rho_{k,m}(t) = 1$ if the resource block $m$ is allocated to user $k$. Otherwise $\rho_{k,m}(t) = 0$. $I_m$ is the interference caused by other users in external services that are using resource block $m$. $B^U$ is the bandwidth of each resource block. $N_0$ is the noise power spectral efficiency.

When multiple users transmit BCI signals to the BS via uplink channels at the same time, the interference between the transmit signals can cause construction errors at the BS due to the noisy channel.
In our work, we consider the packet error rate when transmitting BCI signals of user $k$ as \cite{chen2020}:
\vspace{-0.1cm}
\begin{equation}
\epsilon_k(t) = \sum_{m \in \mathcal{M}} \rho_{k,m} \epsilon_{k, m},
\label{eq:epsilon}
\end{equation}
where $\epsilon_{k,m} = 1 - \text{exp}\Big(- \frac{z \sigma_U^2}{p_k h_k(t)}\Big)$ is the packet error rate over resource block $m$ with $z$ being a waterfall threshold \cite{chen2020}. 
Given the packet errors, the received BCI signals at the BS can contain noise, e.g., Gaussian, or distortion. From hereafter, we denote the noisy BCI signals received at the BS as $\mathbf{\hat{e}}(t)$ to differentiate  from the raw BCI signals as defined in (\ref{eq:bci-signal-vector}). 

For the downlink channel, the downlink data rate achieved by the BS is calculated by:
\vspace{-0.2cm}
\begin{equation}
r_k^D(t) = B^D \log_2 \Big(1 + \frac{P_B h_{k}(t)}{I_D + B^D N_0}\Big),
\label{eq:downlink-rate}
\end{equation}
where $P_B$ is the transmit power of the BS, $I_D$ an $B_D$ are interference and downlink bandwidth, respectively.
Unlike the uplink transmission, the broadcast downlink transmission can significantly reduce packet errors. Therefore, we assume that the packet error rate in the downlink transmission is negligible compared with the uplink.

\vspace{-0.2cm}
\subsection{BCI Predictor}
\begin{figure}[t]
\centering
\includegraphics[width=1.0\linewidth]{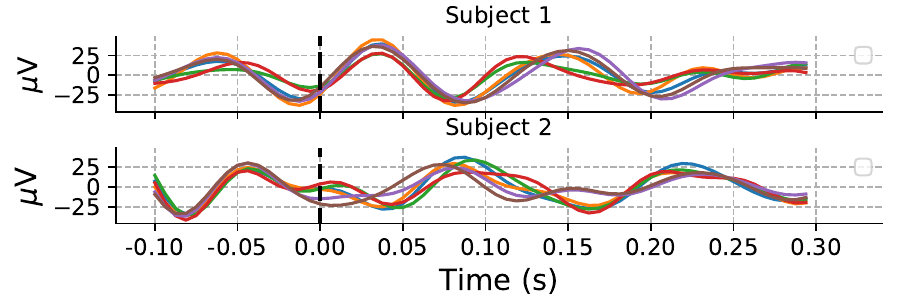}
\caption{An example of EEG signals recorded from two subjects (participants) in dataset \cite{goldberger2000} responding to the same experimental condition. 
The instructions to subjects are placed at the time 0 (marked by the vertical dashed line). 
}
\label{fig:eeg-example}
\end{figure}

We consider a BCI predictor at the BS, denoted by $\phi$, to be a binary indicator (0 or 1) if the predicted output, e.g., predicted hands/feet movement, matches the given labels, denoted by $\mathbf{l}(t)$. In particular, $\phi\left(\mathbf{\hat{e}}(t), \mathbf{l}(t)\right) = 1$ if the prediction is correct. Otherwise $\phi\left(\mathbf{\hat{e}}(t), \mathbf{l}(t)\right) = 0$.
The goal of the BS is to minimize the loss of false detections for the predictor $\phi$ given the collected BCI signals. Formally, we define the loss of the predictor $\phi$ by a cross-entropy loss as follows \cite{sutton2018}:
\vspace{-0.2cm}
\begin{equation}
L_{\phi}\Big(\mathbf{\hat{e}}(t), \mathbf{l}(t)\Big) = -\sum_{c=1}^C \phi\Big(\mathbf{\hat{e}}(t), \mathbf{l}(t)\Big) \log(\varrho_c),
\label{eq:cross-entropy-loss}
\end{equation}
where $C$ is the number of possible actions and $\varrho_c$ is the predicted probability of actions $c$, e.g., moving hands/feet.
In our work, EEG signals are used as BCI signals. However, the extension beyond EEG, e.g., electrocardiogram (ECG) or electromyogram (EMG), is straightforward.
In Fig.~\ref{fig:eeg-example}, we illustrate the EEG signals from two different subjects in dataset \cite{goldberger2000} responding to the same instruction in the experiment, i.e., imagining moving their hands. The goal of the BCI predictor is to accurately predict the action of the user given a fragment of continuous BCI signals as shown in Fig.~\ref{fig:eeg-example}. Our considered setting is even more challenging than conventional BCI settings because our BCI signals contain distortion caused by the noisy wireless environment and the number of resource blocks allocated by the BS.
In the following, we design a QoE model to capture the impacts of the noisy BCI signals on the system performance. 
\vspace{-0.2cm}
\subsection{QoE Model and Problem Formulation}
We consider the QoE of user $k$, denoted by $Q_k$, as a combination of (i) round-trip VR delay and (ii) the prediction accuracy for the actions. Therefore, we define $Q_k$ as follows:
\vspace{-0.3cm}
\begin{multline}
Q_k(\boldsymbol{\rho}, \mathbf{p}, \boldsymbol{\tau}, \phi) = \frac{1}{T} \sum_{t=1}^{T} \Big( \eta_1 \varphi \big(D_k(t), D_{max}\big) + \\ \eta_2 \phi\big(\mathbf{\hat{e}}(t), \mathbf{l}(t)\big) \Big),
\label{eq:qoe-calculation}
\end{multline}
where $\eta_1$ and $\eta_2$ are the positive weight factors; and $T$ is the time horizon. $\varphi(\cdot)$ is also a binary indicator which is defined as follows:
\vspace{-0.2cm}
\begin{equation}
\varphi \Big(D_k(t), D_{max}\Big) = 
    \begin{cases}
      1 & \text{if $D_k(t) \leq D_{max}$,} \\
      0 & \text{otherwise,}
    \end{cases}   
\end{equation}
where $D_{max}$ is the maximum allowed round-trip VR delay for user $k$.
The definition of QoE above reflects the importance of the VR delay and the prediction accuracy of the BCI predictor. The relative importance between the two factors depends on the weighting factors $\eta_1$ and $\eta_2$. For example, in the systems that require highly accurate BCI predictions such as imagined speech communication \cite{lee2022}, the value of $\eta_2$ can be increased. 

In this paper, we consider the optimization problem that maximizes the average QoE of users, given the following constraints: (i) power at the BS and user headsets, (ii) wireless channels, and (iii) computational capability of the BS.
Formally, our optimization problem is defined as follows:
\vspace{-0.2cm}
\begin{subequations}
\label{eq:min-latency}
\begin{align}
\max_{\boldsymbol{\rho}, \mathbf{p}, \boldsymbol{\tau}, \phi} \quad & \frac{1}{K} \sum_{k \in \mathcal{K}} Q_{k}(\boldsymbol{\rho}, \mathbf{p}, \boldsymbol{\tau}, \phi), \\
\textrm{s.t.} \quad & \sum_{k \in \mathcal{K}} \rho_{k,m}(t) = 1, \rho_{k,m}(t) \geq 0, \\
\quad & 0 \leq p_{k}(t) \leq P_{max}, \\
\quad & \sum_{k \in \mathcal{K}}\tau_k(t) \leq 1,  \tau_k \geq 0, \\
\quad & \phi\big(\mathbf{\hat{e}}(t), \mathbf{l}(t)\big) \in \{0, 1\},
\end{align}
\label{eq:qoe-maximization}
\end{subequations}
where $P_{max}$ is the maximum transmission power of the integrated VR-BCI headsets. $\boldsymbol{\rho} = \{\rho_{k, m}(t); \forall k\in \mathcal{K}, \forall m\in \mathcal{M}\}$ is the resource block allocation vector, $\boldsymbol{\tau} = \{\tau_{k}(t); \forall k\in \mathcal{K}\}$ is the computing resource allocation vector, and $\mathbf{p} = \{p_k(t); \forall k \in \mathcal{K}\}$ is the power allocation vector.
In our optimization problem, (\ref{eq:qoe-maximization}b) are the constraints for radio resource block allocation, (\ref{eq:qoe-maximization}c) is the constraint for the transmit power, (\ref{eq:qoe-maximization}d) are the constraints for the computing resource allocation at the BS, and (\ref{eq:qoe-maximization}e) is the BCI predictor constraint.
Note that the maximization of $Q_k$ in (\ref{eq:qoe-maximization}) results in reducing the round-trip VR delay $D_k(t)$ and reducing the BCI prediction loss $L_{\phi}$. 
Our considered problem involves not only a classification problem, i.e., prediction on BCI signals in (\ref{eq:cross-entropy-loss}), but also a decision-making problem, i.e., resource allocation problem in (\ref{eq:processing-delay}) and (\ref{eq:uplink-rate}). 
Therefore, current approaches in BCI classification settings in \cite{lee2022}, and \cite{fernandes2016} can not be directly applied.
In the next section, we propose a novel hybrid learning algorithm to tackle this problem.
 
\section{Proposed Hybrid Learning Algorithm}

\label{sec:hybrid-learning}
\begin{figure}[t]
\centering
\includegraphics[width=1.0\linewidth]{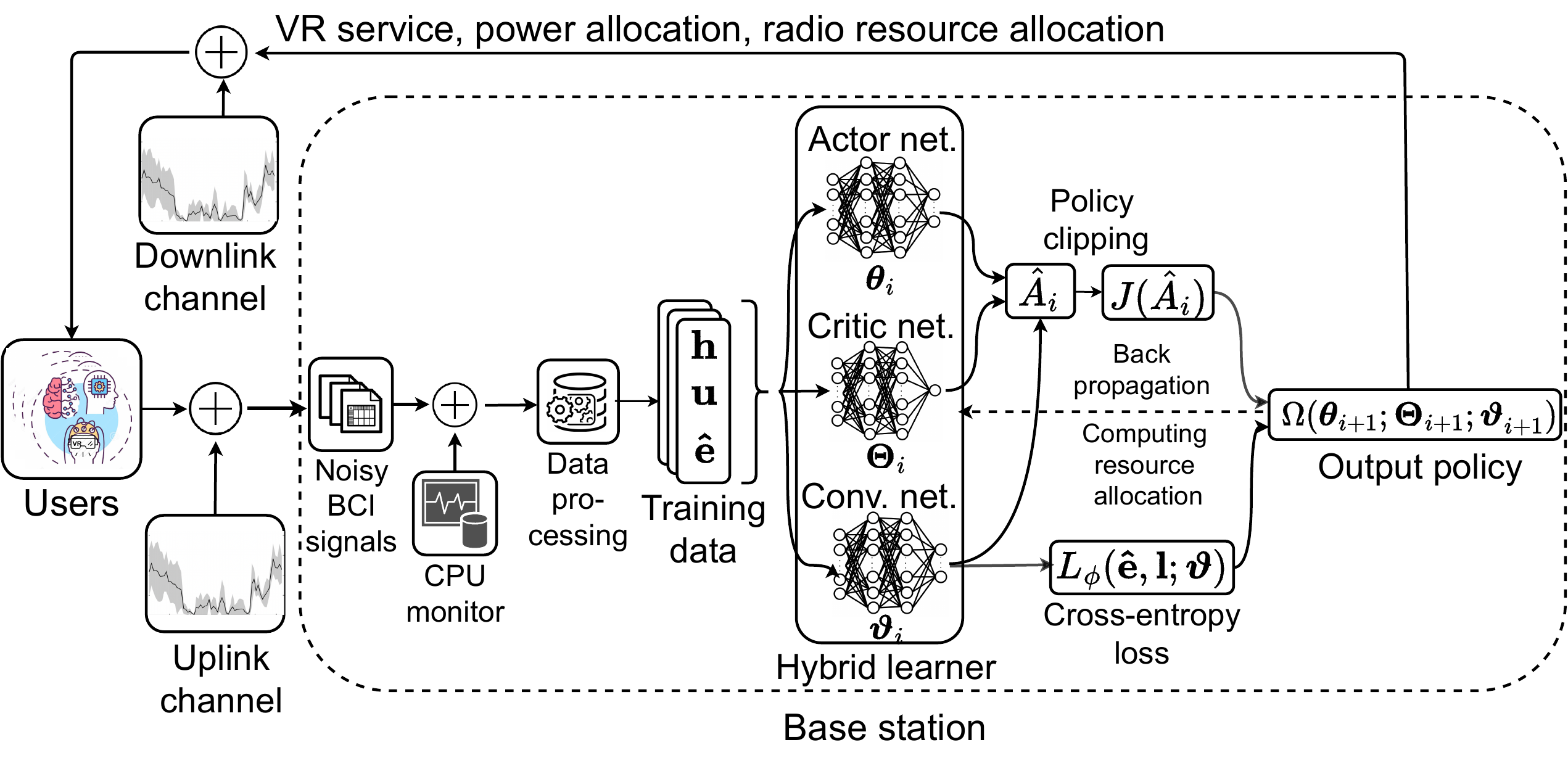}
\caption{Training process for the proposed Hybrid learner.}
\label{fig:hybrid-learner}
\end{figure}

We propose a Hybrid learner which is illustrated in Fig.~\ref{fig:hybrid-learner}. 
Our Hybrid learner consists of three deep neural networks that are (i) an actor network, (ii) a critic network, and (iii) a convolutional network.
The inputs for training the deep neural networks are empirical data from the BCI signals, the wireless channel state, and the computing load of the BS. The output of the proposed algorithm is the policy to jointly allocate power for the users' headsets, allocate radio resources for the uplink channels, and predict the actions of the users based on the BCI signals.
Let $\boldsymbol{\theta}$, $\boldsymbol{\Theta}$, and $\boldsymbol{\varphi}$ denote the parameters, i.e., weights and biases, of the actor network, critic network, and convolutional network, respectively. 
Our proposed  training process for the Hybrid learner is illustrated in Algorithm 1. 

The parameters for deep neural networks are first initialized randomly (line 1 in Algorithm 1).
At each training iteration $i$, the Hybrid learner first collects a set of trajectories $\mathcal{D}_i$ in (\ref{eq:trajectory}) by running current policy $\Omega(\boldsymbol{\theta}_i, \boldsymbol{\Theta}_i, \boldsymbol{\varphi}_i)$ for $O$ time steps. 
The trajectories $\mathcal{D}_i$ contain three main parts that are (i) the observation from the environment, (ii) the action taken of the BS based on the observation from the environment, and (iii) QoE feedback from $K$ users (line 3).
The observation from the environment is a tuple of three states that are channel state $\mathbf{h}$, computing load of the BS $\mathbf{u}$, and BCI signals from users $\mathbf{\hat{e}}$.
The action of the BS is a tuple of four parts that are the radio resource block allocation vector $\boldsymbol{\rho}$, the power allocation vector $\mathbf{p}$, the computing resource allocation vector $\boldsymbol{\tau}$, and the output of the BCI predictor $\phi$.
Based on the collected trajectories, the objective functions for updating the deep neural networks are calculated as follows.
The advantage estimator $\hat{A}_i$ is defined in (\ref{eq:gae-lambda}) \cite{schulman2017} where $\lambda$ is the actor-critic tradeoff parameter and $\delta_o$ is the temporal-difference error which is defined by:
\vspace{-0.1cm}
\begin{equation}
 \delta_o = \frac{1}{K} \sum_{k \in \mathcal{K}}Q_{k,o} + \gamma V(\mathbf{h}_{o+1}, \mathbf{u}_{o+1}, \mathbf{\hat{e}}_{o+1}) - V(\mathbf{h}_{o}, \mathbf{u}_{o}, \mathbf{\hat{e}}_{o}),
 \end{equation}
where $\gamma \in (0, 1)$ is the discount factor and $V(\cdot)$ is the value function of the given observation, i.e., the output of the critic network.
Once the advantage estimator is obtained, the decision-making objective can be calculated by $J(\hat{A}_i)$ as defined in (\ref{eq:decision-making-objective}). In the calculation of $J(\hat{A}_i)$, we adopt a policy-clipping technique from \cite{schulman2017}. In particular, the policy clipping function $g(\varepsilon, \hat{A}_i)$ is:
\begin{equation}
g(\varepsilon, \hat{A}_i)= \begin{cases}(1+\varepsilon) \hat{A}_i, & \text{if } \hat{A}_i \geq 0, \\ (1-\varepsilon) \hat{A}_i, & \text{if } \hat{A}_i < 0.\end{cases}
\end{equation}

\begin{algorithm}[h]
\setstretch{0.5}
\caption{Hybrid Learning Algorithm}
 \textbf{Input}: 
 Initialize $\boldsymbol{\theta}_0$, $\boldsymbol{\Theta}_0$ and $\boldsymbol{\varphi}_0$ at random. \\
 \For{\text{i = 0, 1, 2}, $\ldots$}{
  Collect a set of trajectories $\mathcal{D}_i$:
  \vspace{-0.2cm}
  \begin{multline}
  \mathcal{D}_i = \Big\{\big(\mathbf{h}_o, \mathbf{u}_o, \mathbf{\hat{e}}_o\big), \big(\boldsymbol{\rho}_o, \boldsymbol{p}_o, \boldsymbol{\tau}_o, \phi_o\big), 
  \\ \big(Q_{1,o}, Q_{2,o}, \ldots, Q_{K,o}\big)\Big\}\Big|_{o=1}^{O}.
  \label{eq:trajectory}
  \end{multline} \\
   Compute advantage estimator function $\hat{A}_i$ over $\mathcal{D}_i$:
   \vspace{-0.2cm}
   \begin{equation}
 \hat{A}_i = \sum_{o=1}^{O}(\gamma \lambda)^o \delta_{o}.
 \label{eq:gae-lambda}
  \end{equation} \\
  Calculate the decision-making objective:
  \begin{equation}
    \vspace{-0.1cm}
  J(\hat{A}_i) = \min\Big(\frac{\Omega(\boldsymbol{\theta}_i, \boldsymbol{\Theta}_i, \boldsymbol{\varphi}_i)}{\Omega(\boldsymbol{\theta}_{i-1}, \boldsymbol{\Theta}_{i-1}, \boldsymbol{\varphi}_{i-1})} \hat{A}_i, g(\varepsilon, \hat{A}_i)\Big).
  \label{eq:decision-making-objective}
  \end{equation} \\
  Calculate $L_{\phi}(\mathbf{\hat{e}}, \mathbf{l}; \boldsymbol{\varphi})$ as defined in (\ref{eq:cross-entropy-loss}). \\
  Update the actor network as follows:
  \begin{equation}
  \boldsymbol{\theta}_{i+1} = \boldsymbol{\theta}_{i} + \alpha_a \nabla J(\hat{A}_i).
  \label{eq:actor-net-update}
  \end{equation} \\
  Update the critic network as follows:
  \begin{equation}
  \boldsymbol{\Theta}_{i+1} = \boldsymbol{\Theta}_{i} - \alpha_c \nabla L_c(\boldsymbol{\Theta}).
  \label{eq:critic-net-update}
  \end{equation} \\
  Update the convolutional network as follows:
  \begin{equation}
  \boldsymbol{\varphi}_{i+1} = \boldsymbol{\varphi}_{i} - \alpha_n \nabla L_{\phi}(\mathbf{\hat{e}}, \mathbf{l}; \boldsymbol{\varphi}).
  \label{eq:convo-net-update}
  \end{equation}
 }
\label{algo:hybrid-learner}
\end{algorithm}

With the policy clipping function, the gradient step update is expected not to exceed certain thresholds so that the training  is more stable.
Next, the classification loss $L_{\phi}(\mathbf{\hat{e}}, \mathbf{l}; \boldsymbol{\varphi})$ is calculated based on (\ref{eq:cross-entropy-loss}) with convolutional network (line 6).
The deep neural networks' parameters are finally updated as follows.
The actor network's parameters are updated in (\ref{eq:actor-net-update}) (line 7) where $\alpha_c$ is the learning step size of the actor network and $\nabla$ is the gradient of the function which can be calculated with stochastic gradient decent/ascent algorithms. In our paper, we use Adam as the optimizer for all the deep neural networks.
The critic network's parameters are updated in (\ref{eq:critic-net-update})   (line 8) where $\alpha_c$ is the learning step size and $L_c$ is the critic loss which is defined by:
$ L_c(\boldsymbol{\Theta}) = \Big(V(\mathbf{h}_{o}, \mathbf{u}_{o}, \mathbf{\hat{e}}_{o}) - \frac{1}{K}\sum_{k \in \mathcal{K}}Q_{k,o}\Big)^2.$
Finally, the convolutional network's parameters are updated in (\ref{eq:convo-net-update}) (line 9) where $\alpha_n$ is the learning step size.

\section{Performance Evaluation}
\label{sec:performance-evaluation}
\subsection{Data Processing}
\begin{figure*}[t]
	\centering
	\begin{subfigure}[b]{0.3\linewidth}
		\centering
		\includegraphics[width=1.0\linewidth]{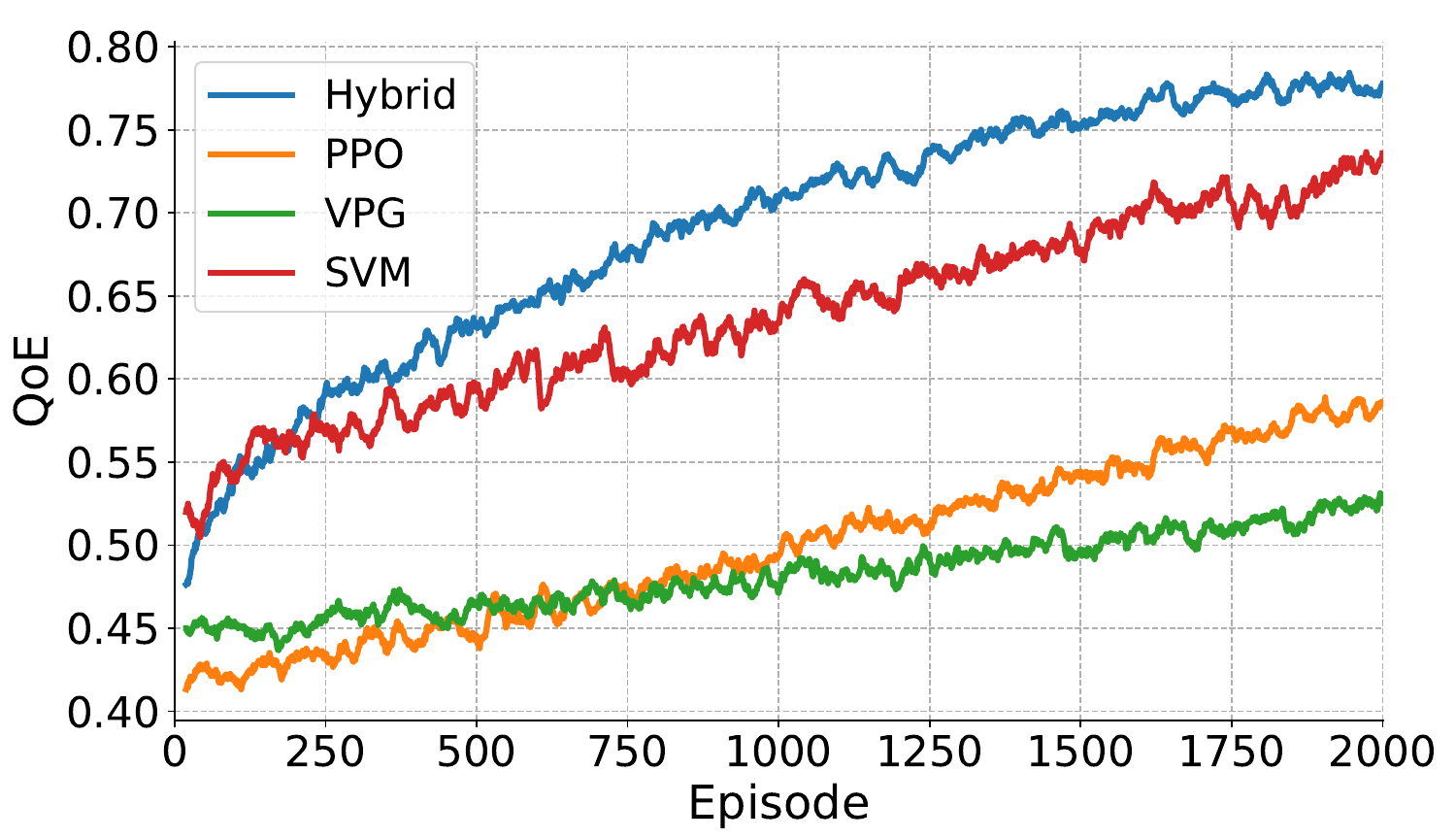}
	\end{subfigure}%
	~ 
	\begin{subfigure}[b]{0.3\linewidth}
		\centering
		\includegraphics[width=1.0\linewidth]{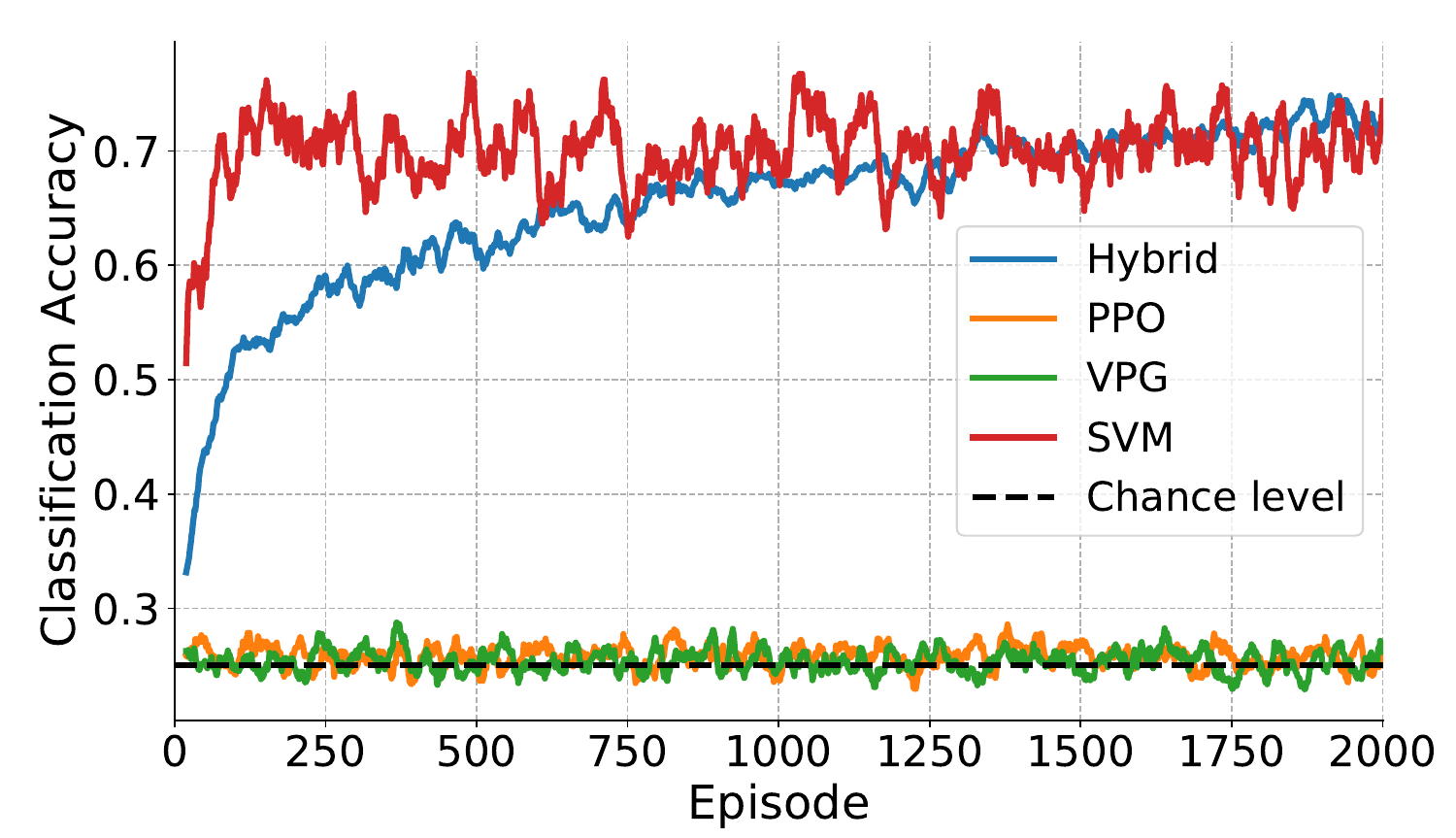}
	\end{subfigure}%
	~
	\begin{subfigure}[b]{0.3\linewidth}
		\centering
		\includegraphics[width=1.0\linewidth]{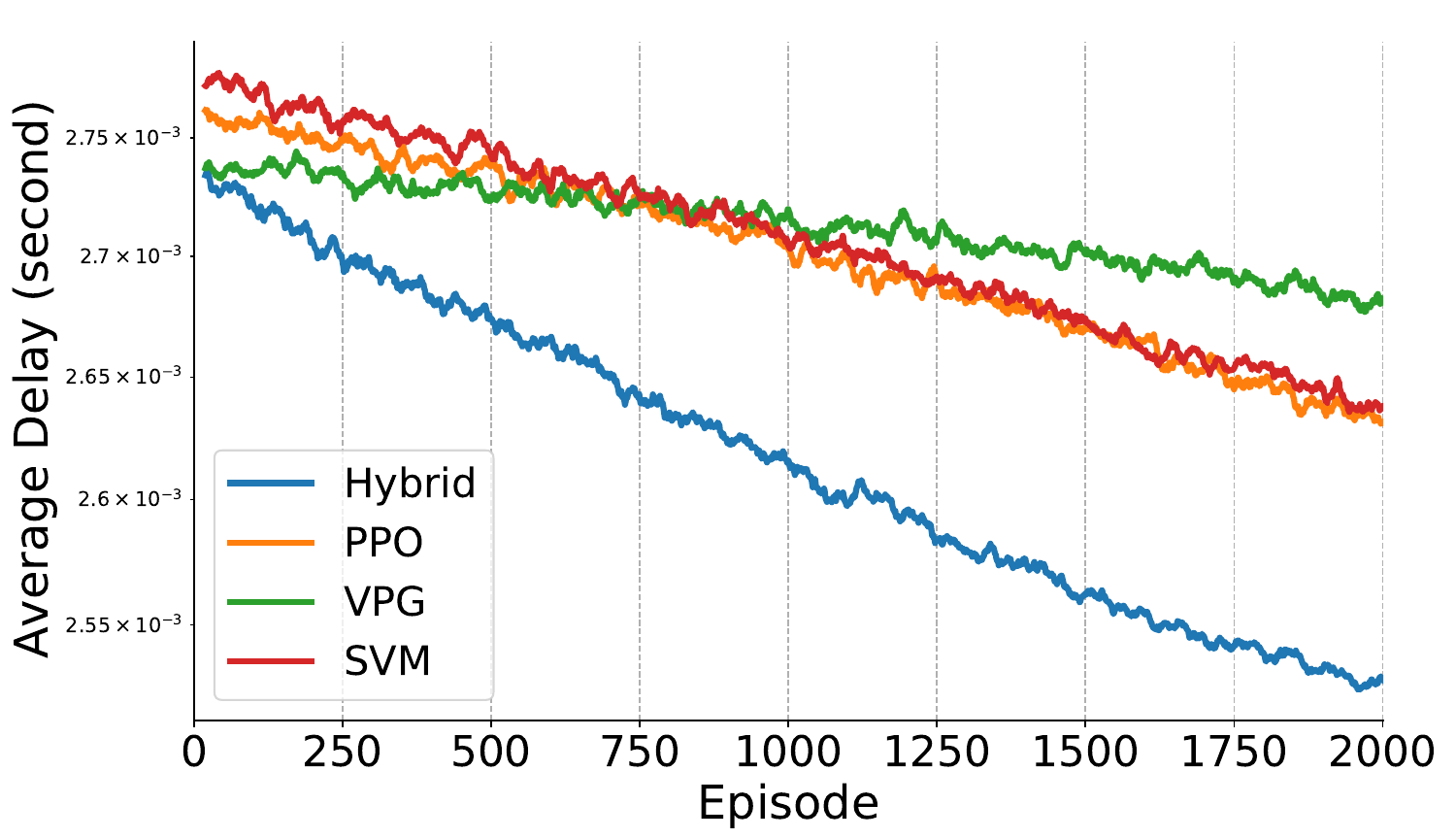}
	\end{subfigure}
	\caption{(a) Normalized QoE, (b) classification accuracy, and (c) average round-trip VR delay values with $(\eta_1, \eta_2) = (1, 1)$.} 
	\label{fig:training-results}
\end{figure*}
We conduct various simulation scenarios to show the efficiency of our proposed approach. In particular, for the BCI classification problem, we use a public dataset from~\cite{goldberger2000}. The dataset contains the experiment results of 109 subjects. Each subject is instructed to do an action per experimental run. The  actions are opening/closing eyes, fists, and feet. 
On each experimental run, the EEG signals are obtained through 64 EEG channels with the BCI2000 system~\cite{schalk2004}. The sampling rate is 160 Hz. In our setting, we consider four different actions, i.e., $C=4$, that are open eyes, close eyes, open fist, and moving feet.
In the default setting, we consider BCI signals from three subjects as illustrated in Fig.~\ref{fig:eeg-example}. 
After processing, the BCI signals of each subject have 255,680 data samples. 
We split the data samples into training and testing datasets with the ratio 80:20. We then use each EEG channel as feature input for the convolutional network of the Hybrid learner. Thus, we have 64 input features and 4 class labels to train with the Hybrid learner. 

For the decision-making problem, i.e., radio and computing resource allocation, we conduct real-time experiments to measure the processing latency at the BS as follows. 
We take panoramic videos from an online source to generate 360-degree views that can be displayed at the user headsets. 
For this, we use the Vue-VR software \cite{ibrahim2022} to pre-render the VR content from the given panoramic videos. We then measure the CPU usage of the local server when running the Vue-VR application. The measured CPU information is used to construct the vector $\mathbf{u}(t)$ in (\ref{eq:cpu-load}).
Our local server is a MacBook Air 2020 with 8GB memory and a 2.3 GHz 8-core CPU.
For the uplink and downlink latency, we use the Rayleigh fading to simulate the dynamics of the time-varying wireless channel. The number of radio resource blocks is set to $M=6$. The number of users is $K=3$. The power of the BS and the user headset are $P_{B}=1$ W and $P_{max} = 0.01$ W.
The uplink and downlink bandwidth are $B^U=1$ MHz and $B^D=20$ MHz \cite{chen2020}. The interference values are $I_m = I_D = -207$ dBm and $N_0 = -174$ dBm.

In comparison with our proposed algorithm, we introduce the following baselines.
(i) \textbf{Proximal Policy Optimization (PPO)} \cite{schulman2017}: PPO is a state-of-the-art reinforcement learning algorithm for decision-making problems. Our Hybrid learner also adopts the actor-critic architecture and policy clipping techniques from PPO to achieve robust performance. We directly use this architecture to learn the QoE defined in (\ref{eq:qoe-calculation}). By maximizing the average QoE, the PPO baseline is expected to reduce the loss $L_{\phi}$ and the round-trip VR delay $D_k$.
(ii) \textbf{Vanilla Policy Gradient (VPG)} \cite{sutton2018}: VPG is a classic policy gradient algorithm for decision-making problems with continuous action values. The VPG baseline also uses the actor-critic architecture. However, the VPG algorithm does not have the embedded advantage estimator function $\hat{A}$ and the policy clipping technique. 
(iii) \textbf{Support-Vector Machine (SVM)} \cite{cortes1995}: SVM is a classic supervised learning algorithm and is a robust benchmark for classification problems. For a fair comparison, we consider the following setting to give SVM certain advantages compared with our proposed algorithm. We train the SVM with training data that are collectively fed into the input of the SVM. In other words, all the training data is stored and reused at the BS. We observe that this training method can significantly boost the performance of the SVM. Otherwise, when we apply the same training method as our proposed algorithm, i.e., the training data is removed after feeding into the deep neural networks, the performance of the SVM is significantly decreased.

\subsection{Simulation Results}
We first illustrate the training performance of the proposed algorithm and the baselines in Fig.~\ref{fig:training-results}. In Fig.~\ref{fig:training-results}, we can observe the increase in QoE values of all the algorithms during 2,000 training episodes. These results imply that all algorithms can learn a good policy given the dynamics of the environment. Note that we consider 2,000 training episodes and terminate the training after 2,000 training episodes to prevent over-fitting. After training, the trained models will be used to evaluate the performance of the learning approaches.
In Fig.~\ref{fig:training-results}(b), we can observe that the proposed Hybrid learner can obtain highly accurate predictions on BCI signals. The SVM baseline also achieves similar performance with the Hybrid learner. Specifically, the SVM baseline learning speed is higher than that of the Hybrid learner because the training data is stored and reused at the BS when we train the SVM baseline. However, the SVM baseline with higher demand for the amount of input data can only converge to the accuracy that is similar to the Hybrid learner.  
Unlike the Hybrid learner and SVM, the performance of the PPO and VPG baselines shows that they are not effective to deal with the classification problem with only reinforcement learning design. With the number of class labels being $C=4$, the prediction accuracy of the PPO and VPG baselines are just approximate the chance level, i.e., $25\%$. 
In Fig.~\ref{fig:training-results}(c), we can observe that all algorithms can reduce the round-trip VR delay. Thanks to reinforcement learning techniques, all algorithms can learn the dynamics of radio and computing resources of the system. We can observe that our proposed algorithm achieves lower round-trip VR delay because of two main reasons.
First, they utilize actor-critic architecture and policy clipping techniques of PPO. Second, our new design in forwarding the losses through the actor-critic networks and convolutional network enables the better realization of the Hybrid learner. As a result, the Hybrid learner can distinguish between two distinct learning goals which are BCI classification and radio/computing resource allocation, and thus facilitating the training process.

\begin{figure}[t]
	\centering
	\begin{subfigure}[b]{0.48\linewidth}
		\centering
		\includegraphics[width=1.0\linewidth]{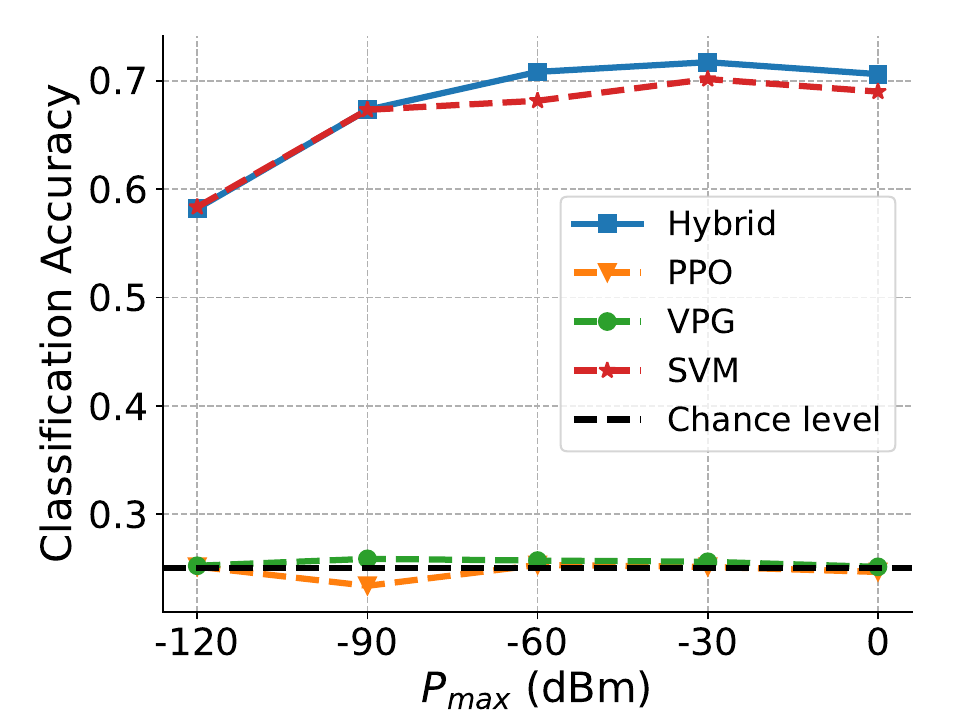}
	\end{subfigure}%
	~
	\begin{subfigure}[b]{0.48\linewidth}
		\centering
		\includegraphics[width=1.0\linewidth]{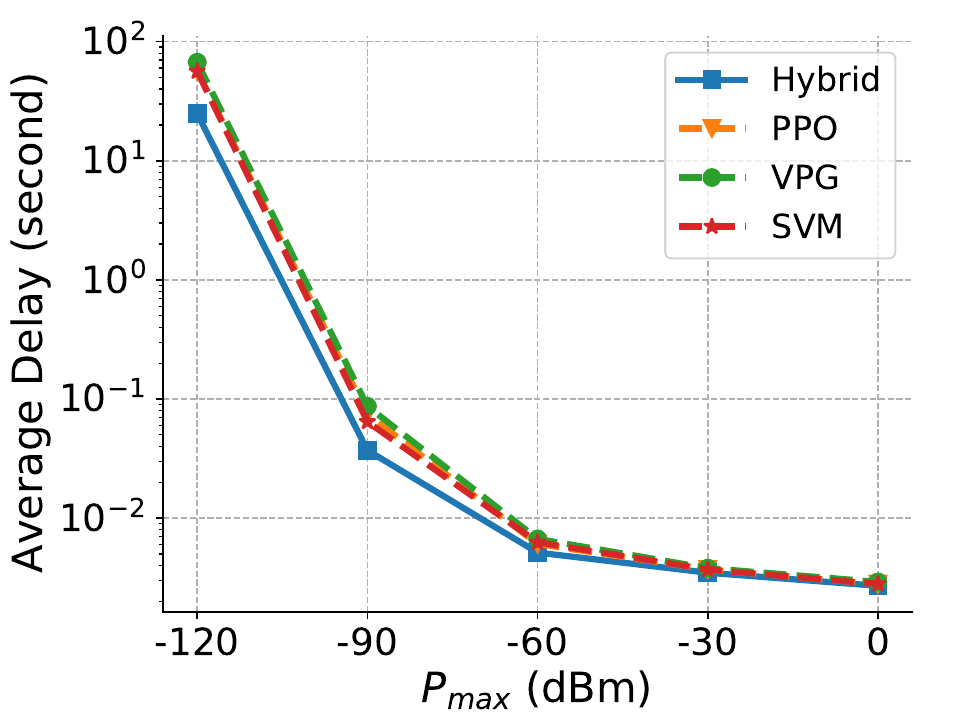}
	\end{subfigure}
	\caption{(a) BCI classification accuracy and (b) round-trip VR delay of the algorithms with testing data when the power capacity varies.} 
	\label{fig:power-varies}
\end{figure}

Next, we vary the maximum power value at the headsets of users, i.e., $P_{max}$, to evaluate the impacts of the power allocation on the system performance. In Fig.~\ref{fig:power-varies}(a), we can observe that when the maximum power of the headsets is close to the noise level, the accuracy of the prediction decreases. The reason is that with the low level of power allocated to the radio resource blocks, the signal-to-interference plus noise ratio (SINR) at the BS may significantly decrease, resulting in the high packet error rate $\epsilon_k$ in (\ref{eq:epsilon}). Our proposed algorithm shows good performance under all the considered scenarios. Similar to the results from Fig.~\ref{fig:training-results}(b), the classification accuracy obtained by the SVM baseline is almost similar to that of the Hybrid learner and is much higher than those of the PPO and VPG baselines. 
In Fig.~\ref{fig:power-varies}(b), we can observe that the increase of power results in the decrease of the round-trip VR delay. The latency obtained by our proposed algorithm is lower than those of the baseline algorithms in most of the scenarios.

\begin{figure}[t]
	\centering
	\begin{subfigure}[b]{0.48\linewidth}
		\centering
		\includegraphics[width=1.0\linewidth]{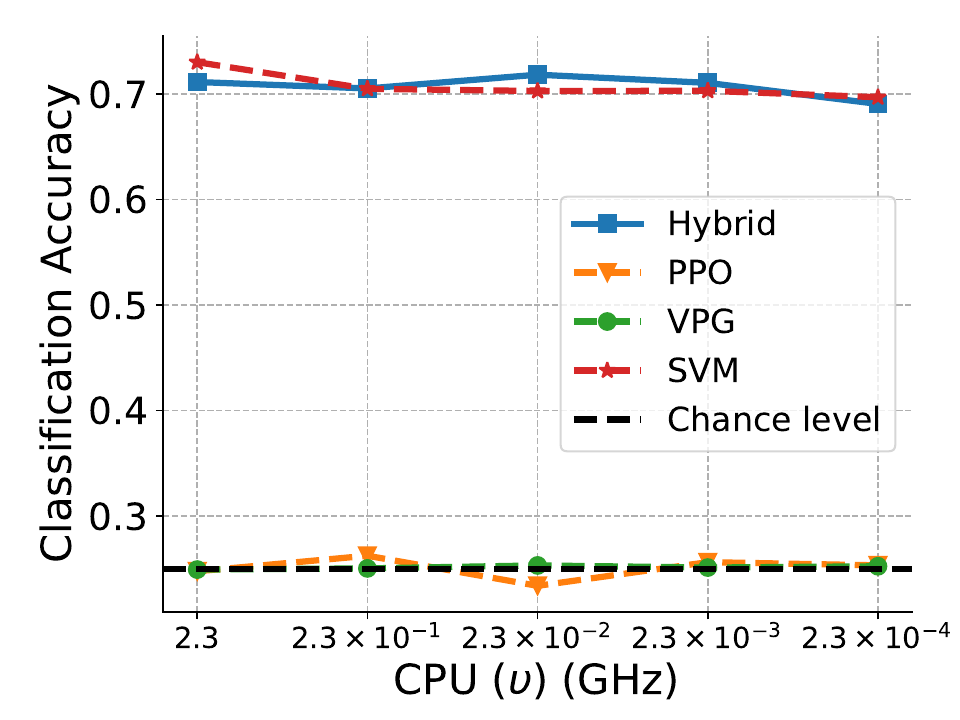}
	\end{subfigure}%
	~
	\begin{subfigure}[b]{0.48\linewidth}
		\centering
		\includegraphics[width=1.0\linewidth]{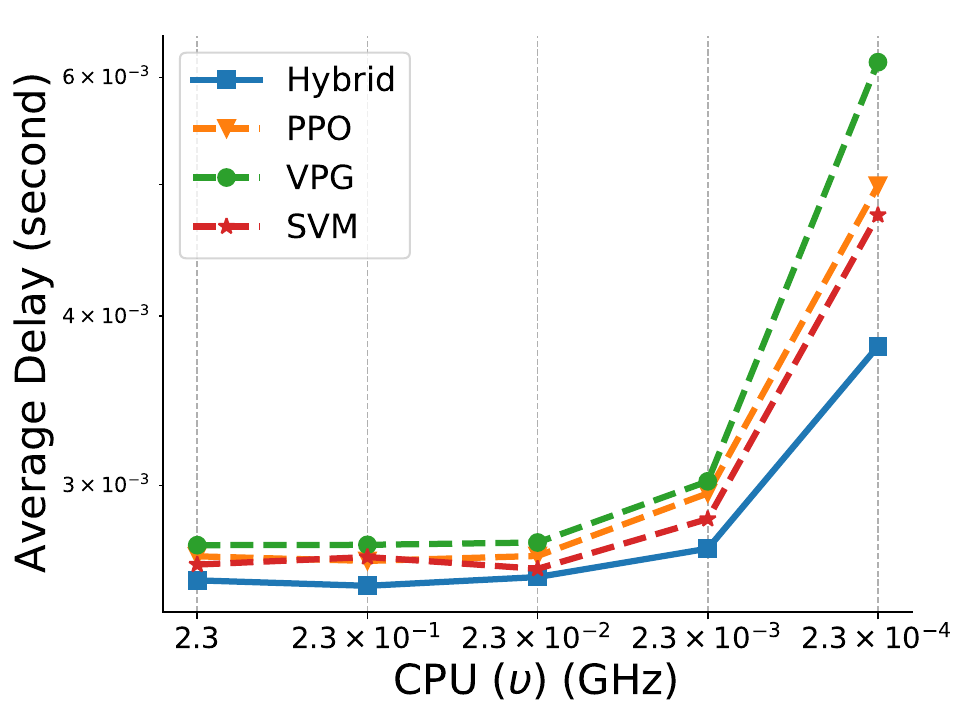}
	\end{subfigure}
	\caption{(a) Classification accuracy and (b) round-trip VR delay of the algorithms with testing data when the CPU capacity of the BS varies.} 
		\label{fig:cpu-varies}
\end{figure}
\vspace{-0.1cm}

Finally, we evaluate the impacts of the computing capacity of the BS on the system performance. In Fig. \ref{fig:cpu-varies}(a), it can be observed that the decrease in the CPU capacity of the BS does not have an impact on the classification accuracy. These results imply that with the limited CPU capacity, our proposed algorithm with deep neural networks can achieve good predictions on the BCI signals. Unlike our proposed algorithm with advanced architecture designs, the PPO and VPG baselines only obtain the classification accuracy values at the chance level, i.e., $25\%$. In Fig.~\ref{fig:cpu-varies}(b), we can observe that the decrease in CPU capacity results in the increase of the round-trip VR delay. The reason is that with lower CPU capacity, the BS takes a longer time to pre-precess VR content for the users.
\section{Conclusion}
In this paper, we have proposed a novel framework to facilitate the creation of intelligent digital avatars for Metaverse users. 
Specifically, our proposed hybrid training algorithm can accurately predict user behaviors by analyzing their brain signals. Furthermore, our proposed algorithm can optimally allocate radio and computing resources to the users so that the end-to-end latency of the system can be minimized. 
Simulation results have shown that our proposed framework with a hybrid learning algorithm outperforms the current state-of-the-art deep reinforcement learning algorithm.

\end{document}